\documentclass[submission,copyright,creativecommons,fleqn]{eptcs}
\providecommand{\event}{ICE 2011} 
\usepackage{breakurl}             
\usepackage{color}
\usepackage{listings}
\usepackage{amsmath,amsthm,amssymb}
\usepackage{mathpartir}
\usepackage{upgreek}
\usepackage{mathbbol}
\usepackage{stmaryrd}

\definecolor{lstbackground}{gray}{0.9}
\definecolor{lstdarkblue}{rgb}{0,0,0.5}

\lstdefinestyle{SINGSHARP}
  {language=C,
   mathescape=true,
   print=true,
   columns=fullflexible,
   basicstyle=\ttfamily,
   identifierstyle=,
   keywordstyle=\bfseries\sffamily,
   morekeywords={receive,send,delete,expose,null,claims,dispose,contract,initial,final,state,open,close,imp,exp,new},
   morecomment=[s][\color{lstdarkblue}]{(*}{*)},
   commentstyle=\color{lstdarkblue},
    literate= 
    {->}{$\to$}1
    {\ ->}{ $\to$}3
    {\ ->\ }{ $\to$ }3
  }

\lstnewenvironment{SingSharp}
  {\lstset{style=SINGSHARP}}
  {}

\newcommand{\Code}{\lstinline[style=SINGSHARP]}

\newif\iflong
\longtrue
\longfalse

\newif\ifcomments
\commentstrue
\commentsfalse

\newif\ifproofs
\proofstrue
\proofsfalse

\ifcomments
\newcommand{\REVISION}[1]{{\color{magenta}#1}}
\else
\newcommand{\REVISION}[1]{#1}
\fi

\newcommand{\eoe}{\hfill\hbox{$\blacksquare$}}

\definecolor{lucared}{rgb}{0.5,0,0}
\definecolor{lucalert}{rgb}{0.8,0,0}
\definecolor{lucagreen}{rgb}{0,0.3,0}

\newcommand{\rulename}[1]{\text{{\sc(#1)}}}

\newcommand{\natset}{\mathbb{N}}

\newcommand{\Fsub}{$F_{{<}{:}}$}
\newcommand{\Sharp}[1]{\textsf{#1}{$\sharp$}}
\newcommand{\CSharp}{\Sharp{C}}
\newcommand{\MiniSing}{\Sharp{CoreSing}}
\newcommand{\PolySing}{\Sharp{PolySing}}
\newcommand{\Sing}{\Sharp{Sing}}

\newcommand{\Process}{\ProcessP}
\newcommand{\ProcessP}{P}
\newcommand{\ProcessQ}{Q}
\newcommand{\ProcessR}{R}

\newcommand{\PointerSet}{\mathtt{Pointers}}
\newcommand{\Pointer}{\PointerA}
\newcommand{\PointerA}{a}
\newcommand{\PointerB}{b}
\newcommand{\PointerC}{c}
\newcommand{\PointerD}{d}
\newcommand{\PointerE}{e}
\newcommand{\PointerF}{f}

\newcommand{\Type}{\TypeT}
\newcommand{\TypeT}{t}
\newcommand{\TypeS}{s}

\newcommand{\SessionType}{\SessionTypeT}
\newcommand{\SessionTypeT}{T}
\newcommand{\SessionTypeS}{S}

\newcommand{\Name}{\NameA}
\newcommand{\NameA}{u}
\newcommand{\NameB}{v}

\newcommand{\VarSet}{\mathtt{Variables}}
\newcommand{\Var}{\VarA}
\newcommand{\VarA}{x}
\newcommand{\VarB}{y}

\newcommand{\RecVar}{\RecVarA}
\newcommand{\RecVarA}{{X}}

\newcommand{\Tag}{\mathtt{m}}

\newcommand{\Memory}{\mu}

\newcommand{\Queue}{q}

\newcommand{\BoundContext}{\Updelta}
\newcommand{\EmptyBoundContext}{{\bullet}}
\newcommand{\RecContext}{\Upsigma}
\newcommand{\Context}{\Upgamma}
\newcommand{\EmptyContext}{\emptyset}

\newcommand{\EmptyMemory}{\emptyset}
\newcommand{\tuple}[1]{(#1)}
\newcommand{\msg}[2]{#1(#2)}
\newcommand{\mtag}[1]{\mathtt{#1}}
\newcommand{\EmptyQueue}{\varepsilon}

\newcommand{\system}[2]{(#1;#2)}

\newcommand{\parop}{\mathbin{|}}
\newcommand{\choice}{\oplus}

\newcommand{\idle}{\textbf{\color{lucagreen}0}}
\newcommand{\primitive}[1]{\text{\sffamily\bfseries#1}}
\newcommand{\openChannel}{\primitive{open}}
\newcommand{\closeChannel}{\primitive{close}}

\newcommand{\send}[3]{#1!#2(#3)}

\newcommand{\rec}{\primitive{rec}~}

\newcommand{\letin}[3]{\mathtt{let}~#1=#2~\mathtt{in}~#3}
\newcommand{\ite}[3]{\mathtt{if}~#1~\mathtt{then}~#2~\mathtt{else}~#3}

\newcommand{\Top}{\mathsf{Top}}

\newcommand{\tvar}{\tvarA}
\newcommand{\tvarA}{\alpha}
\newcommand{\tvarB}{\beta}

\newcommand{\tmsg}[3]{#1\langle#2\rangle(#3)}

\newcommand{\Ref}[1]{{\ast}#1}
\newcommand{\Open}{{\color{lucared}\bullet}}

\newcommand{\tbound}[2]{#1\subt#2}
\newcommand{\trec}{\mu}
\newcommand{\SessionEnd}{{\color{lucagreen}\primitive{end}}}
\newcommand{\InternalChoice}[1]{{!}\{#1\}}
\newcommand{\ExternalChoice}[1]{{?}\{#1\}}

\newcommand{\drel}{\mathcal{D}}
\newcommand{\srel}{\mathcal{S}}
\newcommand{\wrel}{\mathcal{W}}
\newcommand{\red}[1]{\rightarrow_{#1}}
\newcommand{\nred}[1]{\arrownot\red{#1}}
\newcommand{\wred}[1]{\Rightarrow_{#1}}
\newcommand{\eqdef}{\stackrel{\mathrm{def}}{=}}

\newcommand{\subt}{\leqslant}

\newcommand{\wbound}{\mathrel{\downarrow}}
\newcommand{\wbb}[3]{#1\vdash#2\wbound#3}

\renewcommand{\land}{\mathrel{\&}}

\newcommand{\co}[1]{\overline{#1}}
\newcommand{\dom}{\mathsf{dom}}

\newcommand{\subst}[2]{\{#1/#2\}}
\newcommand{\fn}{\mathsf{fn}}
\newcommand{\bn}{\mathsf{bn}}
\newcommand{\tail}{\mathsf{tail}}

\newcommand{\reachable}[2]{\mathsf{reach}(#1,#2)}
\newcommand{\weight}[1]{\|#1\|}
\newcommand{\xweight}[2]{\|#2\|_{#1}}

\title{Polymorphic Endpoint Types for Copyless Message Passing}
\author{Viviana Bono \qquad\qquad Luca Padovani
\institute{Dipartimento di Informatica, Universit\`a di Torino, Italy
\iflong
\fi
}}
\def\titlerunning{Polymorphic Endpoint Types for Copyless Message Passing}
\def\authorrunning{V. Bono, L. Padovani}

\newtheorem{definition}{Definition}[section]

\newtheorem{proposition}{Proposition}[section]
\newtheorem{theorem}{Theorem}[section]

\newtheorem{example}{Example}[section]

\begin{document}
\maketitle

\begin{abstract}
  We present \PolySing{}, a calculus that models process interaction
  based on copyless message passing, in the style of Singularity OS.
  We equip the calculus with a type system that accommodates
  polymorphic endpoint types, which are a variant of polymorphic
  session types, and we show that well-typed processes are free from
  faults, leaks, and communication errors.
  The type system is essentially linear, although linearity alone may
  leave room for scenarios where well-typed processes leak memory. We
  identify a condition on endpoint types that prevents these leaks
  from occurring.
\end{abstract}

\section{Introduction}
\label{sec:introduction}

\newcommand{\foo}{\Code{foo}}

Singularity OS is the prototype of a dependable operating system where
processes share the same address space and interact with each other
solely by message passing over asynchronous channels.
The overhead of communication-based interactions is tamed by copyless
message passing: only \emph{pointers} to messages are physically
transferred from one process to another. Static analysis of
Singularity processes guarantees \emph{process isolation}, namely that
every process can only access memory it owns exclusively.

In \cite{BonoMessaPadovani11} we presented \MiniSing{}, a
formalization of the core features of \Sing{} -- the language used for
the implementation of Singularity OS -- along with a type system
ensuring that well-typed processes are free from communication errors,
memory faults, and memory leaks.
At first sight it might seem that these properties can be trivially
enforced through a linear type system based on session
types~\cite{Honda93,HondaVasconcelosKubo98}. However,
in~\cite{BonoMessaPadovani11} we remarked how linearity alone can be
too restrictive in some contexts and too permissive in others.
To illustrate why linearity can be too restrictive, consider the code
fragment
\begin{equation*}
  \text{\Code{expose (a) \{ send(b, arg, *a); *a := new T(); \}}}
\end{equation*}
which dereferences the pointer \Code{a} and sends \Code{*a} on the
endpoint \Code{b}.  Linearity is violated right after the
\Code{send(arg, b, *a)} command, since \Code{*a} is owned both by the
sender (indirectly, through \Code{a}) as well as by the receiver.
The construct \Code{expose} is used by the \Sing{} compiler to keep
track of memory ownership. In particular, \Sing{} allows pointer
dereferentiation only within \Code{expose} blocks.  The semantics of
an \Code{expose(a)} block is to temporarily transfer the ownership of
\Code{*a} from \Code{a} to the process exposing the pointer. If the
process still owns \Code{*a} at the end of the \Code{expose} block,
the construct is well typed.
In~\cite{BonoMessaPadovani11} we showed that all we need to capture
the static semantics of \Code{expose(a)} blocks is to distinguish
cells with type $\Ref\TypeT$ (whose content, of type $\TypeT$, is
owned by the cell) from cells with type $\Ref\Open$ (whose content is
owned directly by the process). At the beginning of the \Code{expose}
block, \Code{a} is accessed and its type turns from $\Ref\TypeT$ to
$\Ref\Open$; within the block it is possible to (linearly) use
\Code{*a}; at the end of the block, \Code{*a} is assigned with the
pointer to a newly allocated object that the process owns, thus
turning \Code{a}'s type from $\Ref\Open$ back to some $\Ref\TypeS$.

An example where linearity can be too permissive is given by the code
fragment
\begin{equation}
\label{micidiale}
  \text{\Code{(e, f) := open(); send(e, arg, f); close(e);}}
\end{equation}
which creates the two endpoints \Code{e} and \Code{f} of a channel,
sends \Code{f} as the argument of an \Code{arg}-tagged message on
\Code{e}, and closes \Code{e}. This code uses \Code{e} and \Code{f}
linearly and is well typed by associating \Code{e} and \Code{f} with
suitable endpoint types $\SessionTypeT$ and $\SessionTypeS$, where
$\SessionTypeT = {!}\mathtt{arg}(\SessionTypeS).\SessionEnd$ and
$\SessionTypeS = \trec\tvar.{?}\mathtt{arg}(\tvar).\SessionEnd$.
Observe that the code fragment \eqref{micidiale} uses $\mathtt{e}$ in
accordance with type $\SessionTypeT$ and that $\SessionTypeT$ and
$\SessionTypeS$ are \emph{dual endpoint types} (they describe
complementary actions).
Yet, the code generates a memory leak: after the \Code{close}
instruction, no reference to \Code{f} is available, therefore the
\Code{arg}-tagged message will never be received and \Code{f} will
never be deallocated.
In~\cite{BonoMessaPadovani11} we have shown that the leak produced by
this code originates from the recursive type $\SessionTypeS$ and can
be avoided by imposing a simple restriction on endpoint types. The
idea is to define a notion of \emph{weight} for endpoint types which
roughly gives the ``depth'' of the message queues in the endpoints
having those types and to restrict endpoint types to those having
finite weight. Then, one can show that the code \eqref{micidiale} is
well typed only if endpoint types with infinite weight are allowed.

In this work we present \PolySing{}, a variant of \MiniSing{} where we
add \emph{bounded polymorphism} to endpoint types, along the lines
of~\cite{Gay08}, while preserving all the properties mentioned
earlier.
For instance, the polymorphic endpoint type
${!}\tmsg\Tag\tvar\tvar.{?}\Tag(\tvar).\SessionEnd$ denotes an
endpoint on which it is possible to send an $\Tag$-tagged message with
an argument of any type, and then receive another $\Tag$-tagged
message whose argument has the same type as that of the first
message.
%
%
It is possible to specify bounds for type variables, like in
${!}\tmsg\Tag{\tbound{\tvar}{\Type}}\tvar.{?}\Tag(\tvar).\SessionEnd$,
to denote that the type variable $\tvar$ ranges over any
\emph{subtype} of $\Type$, and to recover unconstrained polymorphism
by devising a type $\Top$ that is supertype of any other type.

Now, it may come as a surprise that, when polymorphic endpoint types
are allowed, the code fragment \eqref{micidiale} can be declared well
typed without resorting to recursive types by taking $\SessionTypeT =
{!}\tmsg{\mathtt{arg}}{\tvar}{\tvar}.\SessionEnd$ and $\SessionTypeS =
{?}\tmsg{\mathtt{arg}}{\tvar}{\tvar}.\SessionEnd$.  Since the type
$\SessionTypeT$ of $\mathtt{e}$ is polymorphic, $\mathtt{e}$ accepts a
message with an argument of \emph{any} type and, in particular, a
message with argument $\mathtt{f}$.
Fortunately, a smooth extension of the finite-weight restriction we
have introduced in the monomorphic case rules out this problematic
example. The idea is to estimate the weight of type variables by
looking at their bound. In $\SessionTypeT$ and $\SessionTypeS$ above,
the type variable $\tvar$ has no bound (or, to be precise, is bounded
by the top type $\Top$) and therefore is given infinite weight. When
$\tvar$ occurs in a constraint $\tbound\tvar\Type$, we estimate the
weight of $\tvar$ to be the same as the weight of $\Type$. The
estimation relies on a fundamental relationship between weights and
subtyping, whereby the weight of $\TypeS$ is always smaller than or
equal to the weight of $\TypeT$ if $\TypeS \subt \TypeT$. We show that
forbidding the output of messages whose argument has a type with
infinite weight allows us to prove the absence of leaks, together with
the other desired properties.
Our work is the first to formalize polymorphic endpoint types, which
are effectively described as a feature of Singularity
contracts~\cite{SDN5} even though, to the best of our knowledge, they
never made it to the prototype implementation of Singularity.
Also, the availability of polymorphic endpoint types allows us to
encode linear mutable cells. This renders the $\Ref\Type$ and
$\Ref\Open$ types superfluous and makes our model even more essential
with respect to the one presented in~\cite{BonoMessaPadovani11}.

The rest of the paper is organized thus: Section~\ref{sec:example}
presents an example to introduce all the fundamental concepts
(endpoint types, subtyping, bounded polymorphism) of this work; in
Section~\ref{sec:language} we define the syntax and operational
semantics of \PolySing{} and we formalize the notion of well-behaved
process as a process where faults, leaks, and communication errors do
not occur; Section~\ref{sec:type_system} defines the type system for
\PolySing{} and presents a soundness result (well-typed processes are
well behaved). We discuss related work in Section~\ref{sec:related}
and we draw some conclusions in Section~\ref{sec:conclusion}.
Proofs and additional technical material can be found in the long
version, which is available at the authors' home pages.


\section{An example}
\label{sec:example}

\newcommand{\participant}[1]{\textsf{#1}}
\newcommand{\Buyer}{\participant{Buyer}}
\newcommand{\Seller}{\participant{Seller}}
\newcommand{\Broker}{\participant{Broker}}


It has already been observed in the recent literature that session
types can conveniently describe the interface of distributed objects.
For instance, the session type
\[
  \mathtt{SellerT} =
  \trec\tvar.(
    {!}\mathtt{offer}(\mathtt{nat}).{?}\mathtt{response}(\mathtt{nat}).\tvar
    \oplus
    {!}\mathtt{buy}(\mathtt{string}).\SessionEnd
    \oplus
    {!}\mathtt{leave}().\SessionEnd
  )
\]
may be used to describe (part of) the behavior of a {\Seller} object
as it can be used by {\Buyer}. {\Buyer} may make an offer regarding an
item he or she is interested to buy by sending a $\mathtt{offer}$
message and receiving a counteroffer from {\Seller}. After this
exchange of messages, the protocol repeats itself. Alternatively,
{\Buyer} may $\mathtt{buy}$ the item by sending a message that
contains the delivery address, or it may $\mathtt{leave}$ the virtual
shop.
{\Buyer} and {\Seller} are connected by a pair $(\PointerC,
\PointerD)$ of related endpoints: $\PointerC : \mathtt{SellerT}$ is
given to {\Buyer} which can use it according to the protocol
$\mathtt{SellerT}$; $\PointerD : \co{\mathtt{SellerT}}$ is used by
{\Seller} and its type $\co{\mathtt{SellerT}}$ is the dual of
$\mathtt{SellerT}$, where inputs have been replaced by outputs and
vice versa. This guarantees that {\Buyer} and {\Seller} interact
without errors.

Suppose now that we want to implement a {\Broker} to which {\Buyer}
can delegate the bargaining protocol with {\Seller}. We may describe
{\Broker} in some pseudo-language, as follows:
\lstset{numbers=left,numberstyle=\tt\footnotesize}
\begin{SingSharp}
  BROKER($\PointerB$ : $\co{\mathtt{BrokerT}}$) {
    $p_0$ := receive($\PointerB$, price);
    $x$ := receive($\PointerB$, seller);
    $p$ := $p_0$;
    while (better_deal($p_0$, $p$)) {
      send($x$, offer, compute_new_offer($p_0$, $p$));
      $p$ := receive($x$, response);
    }
    send($\PointerB$, price, $p$);
    send($\PointerB$, seller, $x$);
  }
\end{SingSharp}
{\Buyer} and {\Broker} interact by means of another pair $(\PointerA,
\PointerB)$ of related endpoints, $\PointerA : \mathtt{BrokerT}$ in
use by {\Buyer} (not shown here) and $\PointerB :
\co{\mathtt{BrokerT}}$ in use by {\Broker}.
{\Broker} accepts an initial $\mathtt{price}$ $p_0$ and the endpoint
of {\Seller} from {\Buyer} (lines 2--3). Then, it engages the
bargaining protocol with {\Seller} (lines 4--8) until the best deal
$p$ is achieved. Finally, it returns $p$ and the {\Seller}'s endpoint
to {\Buyer} (lines 9--10) so that {\Buyer} can conclude the
interaction with {\Seller} appropriately.

{\Buyer} interacts with {\Broker} on the endpoint $\PointerA$, whose
type can be described by:
\[
  \mathtt{BrokerT} =
  {!}\mathtt{price}(\mathtt{nat}).
  {!}\mathtt{seller}(\mathtt{SellerT}).
  {?}\mathtt{price}(\mathtt{nat}).
  {?}\mathtt{seller}(\mathtt{SellerT}).
  \SessionEnd
\]

Observe that $\mathtt{BrokerT}$ might be considered too precise, since
{\Broker} uses only a strict subset of the functionalities supported
by {\Seller} and described in $\mathtt{SellerT}$. To maximize
reusability, it could be more appropriate to replace the type
$\mathtt{SellerT}$ in $\mathtt{BrokerT}$ with
\[
  \mathtt{BargainT} = \trec\tvar.{!}\mathtt{offer}(\mathtt{nat}).{?}\mathtt{response}(\mathtt{nat}).\tvar
\]
which specifies the minimum set of functionalities of {\Seller} that
{\Broker} actually uses.
{\Buyer} can still delegate an endpoint of type $\mathtt{SellerT}$ to
{\Broker} since $\mathtt{SellerT}$ is a subtype of
$\mathtt{BargainT}$, which we express as the relation
$\mathtt{SellerT} \subt \mathtt{BargainT}$.  This is consistent with
the notion of subtyping in object-oriented languages: if we think of
the message tags in $\mathtt{SellerT}$ as of the methods provided by
{\Seller}, then $\mathtt{BargainT}$ is one possible \emph{interface}
that {\Seller} implements and in fact is the least interface needed by
{\Broker}.
The problem, in this case, is that when an endpoint is delegated,
{\Buyer} can no longer use it (endpoints are linear resources). It is
true that {\Broker} eventually returns {\Seller}'s endpoint to
{\Buyer}, but by that time its type has been widened to
$\mathtt{BargainT}$ and {\Buyer} can no longer access the
functionalities in $\mathtt{SellerT}$ that have been hidden in
$\mathtt{BargainT}$. This lost information can be recovered using
bounded polymorphism and by giving a more precise type to $\PointerA$:
\[
  \mathtt{BrokerT} =
  {!}\mathtt{price}(\mathtt{nat}).
  {!}\mathtt{seller}\langle\tbound{\tvar}{\mathtt{BargainT}}\rangle(\tvar).
  {?}\mathtt{price}(\mathtt{nat}).
  {?}\mathtt{seller}(\tvar).
  \SessionEnd
\]

In this case, the protocol still allows {\Buyer} to send any endpoint
corresponding to {\Seller} whose type conforms to (is a subtype of)
$\mathtt{BargainT}$, but it also specifies that the endpoint
eventually returned to {\Buyer} has the same type as the one sent
earlier.
In general, once an endpoint has been delegated, the delegator loses
access to the endpoint as well as to any information about its type.
This loss of information, which alone justifies the interest for
bounded polymorphism in sequential programming, is likely to occur
much more frequently in our context where many resources are linear.



\section{Language syntax and semantics}
\label{sec:language}

\begin{table}
\caption{\label{tab:syntax} Syntax of \PolySing{} processes and of heap objects.}
\framebox[\textwidth]{
\begin{math}
\displaystyle
\begin{array}{@{}l@{\qquad\qquad}l@{}}
\begin{array}[t]{@{}rcl@{\quad}l@{}}
  \Process & ::= & & \textbf{Process} \\
  &   & \idle & \text{(idle)} \\
  & | & \RecVar & \text{(variable)} \\
  & | & \closeChannel(\Name) & \text{(close channel)} \\
  & | & \openChannel(\Pointer, \Pointer).\Process & \text{(open channel)} \\
  & | & \send\Name\Tag\Name.\Process & \text{(send)} \\
  & | & \sum_{i\in I} \Name?\Tag_i(\Var_i).\Process_i & \text{(receive)} \\
  & | & \Process \choice \Process & \text{(choice)} \\
  & | & \Process \parop \Process & \text{(composition)} \\
  & | & \rec\RecVar.\Process & \text{(recursion)} \\
\end{array}
&
\begin{array}[t]{@{}rcl@{\quad}l@{}}
  \Memory & ::= & & \textbf{Heap} \\
  &   & \EmptyMemory & \text{(empty)} \\
  & | & \Pointer \mapsto [\Pointer, \Queue] & \text{(endpoint)} \\
  & | & \Memory, \Memory & \text{(composition)} \\
  \\
  \Queue & ::= & & \textbf{Queue} \\
  &   & \EmptyQueue & \text{(empty)} \\
  & | & \msg\Tag\Pointer & \text{(message)} \\
  & | & \Queue :: \Queue & \text{(composition)} \\
\end{array}
\end{array}
\end{math}
}
\end{table}

We fix some notation:
we use $\ProcessP$, $\ProcessQ$, $\dots$ to range over processes and
$\PointerA$, $\PointerB$, $\dots$ to range over \emph{heap pointers}
(or simply \emph{pointers}) taken from some infinite set
$\PointerSet$;
we use $\VarA$, $\VarB$, $\dots$ to range over \emph{variables} taken
from some infinite set $\VarSet$ disjoint from $\PointerSet$ and we
let $\NameA$, $\NameB$, $\dots$ range over \emph{names}, which are
elements of $\PointerSet \cup \VarSet$;
finally, we let $\RecVar$ range over \emph{process variables}.

The language of processes, defined by the grammar in
Table~\ref{tab:syntax}, essentially is a monadic pi-calculus equipped
with tag-based message dispatching and primitives for handling
endpoints.
The process $\idle$ is idle and performs no action.
Terms $\rec\RecVar.\Process$ and $\RecVar$ are used for building
recursive processes, as usual.
The process $\Name!\msg\Tag{\NameB}.\Process$ sends a message
$\msg\Tag{\NameB}$ on the endpoint $\Name$ and continues as
$\Process$. A \emph{message} is made of a \emph{tag} $\Tag$ along with
it \emph{parameter} $\NameB$.
The term $\sum_{i\in I} \Name?\Tag_i(\Var_i).\Process_i$ denotes a
process that waits for a message from the endpoint $\Name$. The tag
$\Tag_i$ of the received message determines the continuation
$\Process_i$ where the variable $\Var_i$ is instantiated with the
parameter of the message. We assume that in every such term the
$\Tag_i$'s are pairwise distinct.
Sometimes we will write $\Name?\Tag_1(\Var_1).\Process_1 + \cdots +
\Name?\Tag_n(\Var_n).\Process_n$ in place of $\sum_{i=1}^n
\Name?\Tag_i(\Var_i).\Process_i$.
The term $\openChannel(\PointerA, \PointerB).\Process$ denotes a
process creating a \emph{channel}, represented as two peer endpoints
$\PointerA$ and $\PointerB$.
The process $\closeChannel(\Name)$ closes the endpoint located at
$\Name$.
The processes $\ProcessP \choice \ProcessQ$ and $\ProcessP \parop
\ProcessQ$ are standard and respectively denote the non-deterministic
choice and the parallel composition of $\ProcessP$ and $\ProcessQ$.
The sets of free and bound names of every process $\Process$,
respectively denoted by $\fn(\Process)$ and $\bn(\Process)$, are
standard: the construct $\openChannel(\PointerA, \PointerB).\Process$
binds both $\PointerA$ and $\PointerB$ in $\Process$, while
$\sum_{i\in I} \Name?\Tag_i(\Var_i).\Process_i$ binds $\Var_i$ in
$\Process_i$ for each $i\in I$.
The construct $\rec\RecVar$ is the only binder for process
variables. We adopt the Barendregt convention for both variables and
process variables. 

\begin{example}[linear mutable cell]
\label{ex:cell}
The following process models a linear mutable cell located at
$\PointerC$:
\[
  \mathtt{CELL}(\PointerC) =
  \rec\RecVar.
  (\PointerC?\mathtt{set}(x).\PointerC!\mathtt{get}(x).\RecVar
  +\PointerC?\mathtt{free}().\closeChannel(\PointerC))
\]

The user of the cell interacts with it on the peer endpoint of
$\PointerC$. Initially, the cell is empty and offers to its user the
possibility of setting (with a $\mathtt{set}$-tagged message) the
content of the cell with a pointer $x$, or deallocating (with a
$\mathtt{free}$-tagged message) the cell. In the first case, the cell
transits into a new state where the only possible operation is
retrieving (with a $\mathtt{get}$-tagged message) the content of the
cell. At that point, the cell returns to its original state.
\REVISION{The cell is \emph{linear} in the sense that it allows
  setting its content only if the previous content has been
  retrieved. This cell implementation resembles that of a 1-place
  buffer, but we retain the name ``cell'' for continuity with our
  previous work~\cite{BonoMessaPadovani11}.}
\eoe
\end{example}

\emph{Heaps}, ranged over by $\Memory$, $\dots$, are finite maps from
pointers to heap objects represented as terms defined according to the
syntax in Table~\ref{tab:syntax}:
the heap $\EmptyMemory$ is empty;
the heap $\PointerA \mapsto [\PointerB, \Queue]$ is made of an
endpoint located at $\PointerA$ which is a structure referring to the
peer endpoint $\PointerB$ and containing a \emph{queue} $\Queue$ of
messages waiting to be read from $\PointerA$.
Heap compositions $\Memory,\Memory'$ are defined only when the domains
of the heaps being composed, which we denote by $\dom(\Memory)$ and
$\dom(\Memory')$, are disjoint.
We assume that heaps are equal up to commutativity and associativity
of composition and that $\EmptyMemory$ is neutral for composition.
\emph{Queues}, ranged over by $\Queue$, $\dots$, are finite ordered
sequences of messages $\msg{\Tag_1}{\PointerC_1} :: \cdots ::
\msg{\Tag_n}{\PointerC_n}$. We build queues from the empty queue
$\EmptyQueue$ and concatenation of messages by means of $::$. We
assume that queues are equal up to associativity of $::$ and that
$\EmptyQueue$ is neutral for $::$.

We define the operational semantics of processes as the combination of
a structural congruence relation and a reduction relation.
Structural congruence, denoted by $\equiv$, is the least congruence
relation that includes alpha conversion on bound names and the usual
laws
\[
  \Process \parop \idle \equiv \Process
  \qquad
  \ProcessP \parop \ProcessQ \equiv \ProcessQ \parop \ProcessP
  \qquad
  \ProcessP \parop (\ProcessQ \parop \ProcessR)
  \equiv
  (\ProcessP \parop \ProcessQ) \parop \ProcessR
\]

\begin{table}
\caption{\label{tab:reduction} Reduction of systems.}
\framebox[\textwidth]{
\begin{math}
\displaystyle
\begin{array}{@{}c@{}}
\inferrule[\rulename{R-Open}]{}{
  \system\Memory{\openChannel(\PointerA,\PointerB).\Process}
  \red{}
  \system{
    \Memory,
    \PointerA\mapsto[\PointerB,\EmptyQueue],
    \PointerB\mapsto[\PointerA,\EmptyQueue]
  }{
    \Process
  }
}
\qquad
\inferrule[\rulename{R-Rec}]{}{
  \system\Memory{\rec\RecVar.\Process}
  \red{}
  \system\Memory{\Process\subst{\rec\RecVar.\Process}\RecVar}
}
\\\\
\inferrule[\rulename{R-Choice}]{}{
  \system\Memory{\ProcessP\choice\ProcessQ}
  \red{}
  \system\Memory\ProcessP
}
\qquad
\inferrule[\rulename{R-Send}]{}{
  \system{
    \Memory,
    \PointerA \mapsto [\PointerB, \Queue],
    \PointerB \mapsto [\PointerA, \Queue']
  }{
    \PointerA!\msg\Tag{\PointerC}.\Process
  }
  \red{}
  \system{
    \Memory,
    \PointerA \mapsto [\PointerB, \Queue],
    \PointerB \mapsto [\PointerA, \Queue'::\msg\Tag{\PointerC}]
  }{
    \Process
  }
}
\\\\
\inferrule[\rulename{R-Receive}]{
  k\in I
}{
  \textstyle
  \system{
    \Memory,
    \PointerA \mapsto [\PointerB, \msg{\Tag_k}{\PointerC}::\Queue]
  }{
    \sum_{i\in I} \PointerA?\Tag_i(\Var_i).\Process_i
  }
  \red{}
  \system{
    \Memory,
    \PointerA \mapsto [\PointerB, \Queue]
  }{
    \Process_k\subst{\PointerC}{\Var_k}
  }
}
\qquad
\inferrule[\rulename{R-Par}]{
  \system\Memory\ProcessP \red{} \system{\Memory'}{\ProcessP'}
}{
  \system\Memory{\ProcessP \parop \ProcessQ}
  \red{}
  \system{\Memory'}{\ProcessP' \parop \ProcessQ}
}
\end{array}
\end{math}
}
\end{table}

\PolySing{} processes communicate with each other by means of the
heap. Therefore, the reduction relation defines the transitions of
\emph{systems} instead of processes, where a system is a pair
$\system\Memory\Process$ of a heap $\Memory$ and a process $\Process$.
The reduction relation $\red{}$, inductively defined in
Table~\ref{tab:reduction}, is described in the following paragraph.
\rulename{R-Open} creates a new channel consisting of two fresh
endpoints which refer to each other and have an empty queue.
%
%
\rulename{R-Choice} (and its symmetric, omitted) states that a process
$\ProcessP \choice \ProcessQ$ may autonomously reduce to either
$\ProcessP$ or $\ProcessQ$ leaving the heap
unchanged. 
%
\rulename{R-Send} describes the output of a message
$\msg\Tag{\PointerC}$ on the endpoint $\PointerA$. The message is
enqueued at the right end of $\PointerA$'s peer endpoint queue.
\rulename{R-Receive} describes the input of a message from the
endpoint $\PointerA$. The message at the left end of $\PointerA$'s
queue is removed from the queue, its tag is used for selecting some
branch $k\in I$, and its parameter instantiates the variable
$\Var_k$. 
%
\rulename{R-Par} (and its symmetric, omitted) expresses reductions
under parallel composition. The heap is treated globally, even when it
is only a sub-process to reduce. 
%
\rulename{R-Rec} is the usual unfolding of recursive processes.
Observe that the Barendregt convention makes sure that in the
unfolding $\Process\subst{\rec\RecVar.\Process}{\RecVar}$ of a
recursive process no name occurring free in $\Process$ is accidentally
captured, because $\fn(\Process) \cap \bn(\Process) = \emptyset$.
Not shown in Table~\ref{tab:reduction} is the usual rule
\rulename{R-Struct} describing reductions modulo structural
congruence, which plays an essential role in ensuring that
$\openChannel(\PointerA,\PointerB).\Process$ is never stuck, because
$\PointerA$ and $\PointerB$ can always be alpha converted to some
pointers not occurring in $\dom(\Memory)$.
There is no reduction for $\closeChannel(\PointerA)$
processes. \REVISION{In principle, $\closeChannel(\PointerA)$ should
  deallocate the endpoint located at $\PointerA$ by removing its
  association from the heap. However, since peer endpoints mutually
  refer to each other, removing one endpoint could leave a dangling
  reference from the corresponding peer. Also, it is convenient to
  treat $\closeChannel(\PointerA)$ processes as persistent because, in
  this way, we keep track of the pointers that have been properly
  deallocated. We will see that this information is crucial in the
  definition of well-behaved processes (Definition~\ref{def:wb}).} A
process willing to deallocate a pointer $\PointerA$ and to continue as
$\Process$ afterwards can be modelled as
$\closeChannel(\PointerA) \parop \Process$.
In the following we write $\wred{}$ for the reflexive, transitive
closure of $\red{}$ and we write $\system\Memory\Process \nred{}$ if
there exist no $\Memory'$ and $\Process'$ such that
$\system\Memory\Process \red{} \system{\Memory'}{\Process'}$.

In this work we characterize well-behaved systems as those that are
free from faults, leaks, and communication errors: a \emph{fault} is
an attempt to use a pointer not corresponding to an allocated object
or to use a pointer in some way which is not allowed by the object it
refers to; a \emph{leak} is a region of the heap that some process
allocates and that becomes unreachable because no reference to it is
directly or indirectly available to the processes in the system; a
\emph{communication error} occurs if some process receives a message
of unexpected type.
We conclude this section formalizing these properties.  To do so, we
need to define the reachability of a heap object with respect to a set
of \emph{root} pointers. Intuitively, a process $\Process$ may
directly reach any object located at some pointer in the set
$\fn(\Process)$ (we can think of the pointers in $\fn(\Process)$ as of
the local variables of the process stored in its stack); from these
pointers, the process may reach other heap objects by reading messages
from other endpoints it can reach.

\newcommand{\Pointers}{A}

\newcommand{\xreach}[1][]{\prec_{#1}}
\newcommand{\reach}[1][]{\prec_{#1}^{*}}

\begin{definition}[reachable pointers]
  We say that $\PointerC$ is \emph{reachable} from $\PointerA$ in
  $\Memory$, notation $\PointerC \xreach[\Memory] \PointerA$, if
  $\PointerA \mapsto [\PointerB, \Queue :: \Tag(\PointerC) :: \Queue']
  \in \Memory$.
  We write $\PointerC \reach[\Memory] \PointerA$ for the reflexive,
  transitive closure of $\xreach[\Memory]{}{}$.
  Let $\reachable{\Pointers}{\Memory} = \{ \PointerC \mid
  \exists\PointerA\in\Pointers: \PointerC \reach[\Memory] \PointerA
  \}$.
\end{definition}

We now define well-behaved systems formally.

\begin{definition}[well-behaved process]
\label{def:wb}
We say that $\ProcessP$ is \emph{well behaved} if
$\system\EmptyMemory\ProcessP \wred{} \system\Memory{\ProcessQ}$
implies:
\begin{enumerate}
\item $\dom(\Memory) = \reachable{\fn(\ProcessQ)}\Memory$;
\item $\ProcessQ \equiv \Process_1 \parop \Process_2$ implies
  $\reachable{\fn(\Process_1)}\Memory \cap
  \reachable{\fn(\Process_2)}\Memory = \emptyset$;
\item $\ProcessQ \equiv \Process_1 \parop \Process_2$ and
  $\system\Memory{\Process_1} \nred{}$ where $\Process_1$ does not
  have unguarded parallel compositions imply either $\Process_1 =
  \idle$ or $\Process_1 = \closeChannel(\Pointer)$ or $\Process_1 =
  \sum_{i\in I} \Pointer?\Tag_i(\Var_i).\Process_i$ and $\Pointer
  \mapsto [\PointerB, \EmptyQueue] \in \Memory$.
\end{enumerate}
\end{definition}

\REVISION{In words, a process $\ProcessP$ is well behaved if every
  residual of $\ProcessP$ reachable from a configuration where the
  heap is empty satisfies a number of conditions.}
Conditions~(1) and~(2) guarantee the absence of faults and
leaks. Indeed, condition~(1) states that every pointer to the heap is
reachable by one process, and that every reachable pointer corresponds
to an object allocated on the heap. Condition~(2) states that
processes are isolated, namely that the sets of reachable pointers
corresponding to different processes are disjoint. Since processes of
the form $\closeChannel(\Pointer)$ are persistent, this also
guarantees the absence of faults where a process tries to use an
endpoint that has already been deallocated, or where the same endpoint
is deallocated twice. Condition~(3) guarantees the absence of
communication errors, namely that if $\system\Memory\ProcessQ$ is
stuck (no reduction is possible), then it is because every
non-terminated process in $\ProcessQ$ is waiting for a message on an
endpoint having an empty queue. This configuration corresponds to a
genuine deadlock where every process in some set is waiting for a
message that is to be sent by another process in the same set.
\REVISION{We only consider initial configurations with an empty heap
  for two reasons: first, we take the point of view that initially
  there are no allocated objects; second, since we will need a
  well-typed predicate for heaps and we do not want to verify heap
  well-typedness at runtime, we will make sure that the empty heap is
  trivially well typed.}


\section{Type system}
\label{sec:type_system}

\newcommand{\innervars}{\mathcal{I}}
\newcommand{\outervars}{\mathcal{O}}

\begin{table}
\caption{\label{tab:type_syntax} Syntax of types.}
\framebox[\textwidth]{
\begin{math}
\displaystyle
\begin{array}{@{}l@{\qquad\qquad}l@{}}
\begin{array}[t]{@{}rcl@{\quad}l@{}}
  \SessionType & ::= & & \textbf{Endpoint Type} \\
  &   & \SessionEnd & \text{(termination)} \\
  & | & \tvar & \text{(variable)} \\
  & | & \InternalChoice{\tmsg{\Tag_i}{\tbound{\tvar_i}{\TypeT_i}}{\TypeS_i}.\SessionType_i}_{i\in I} & \text{(internal choice)} \\
  & | & \ExternalChoice{\tmsg{\Tag_i}{\tbound{\tvar_i}{\TypeT_i}}{\TypeS_i}.\SessionType_i}_{i\in I} & \text{(external choice)} \\
  & | & \trec\tvar.\SessionType & \text{(recursive type)} \\
\end{array}
&
\begin{array}[t]{@{}rcl@{\quad}l@{}}
  \TypeT & ::= & & \textbf{Type} \\
  &   & \Top & \text{(top type)} \\
  & | & \SessionType & \text{(endpoint type)} \\
\end{array}
\end{array}
\end{math}
}
\end{table}

We introduce some notation for the type system: we assume an infinite
set of \emph{type variables} ranged over by $\tvarA$, $\tvarB$,
\dots. Types are ranged over by $\TypeT$, $\TypeS$, \dots while
endpoint types are ranged over by $\SessionTypeT$, $\SessionTypeS$,
\dots.
The syntax of types and endpoint types is defined in
Table~\ref{tab:type_syntax}.
An endpoint type describes the behavior of a process with respect to a
particular endpoint. The process may send messages over the endpoint,
receive messages from the endpoint, and deallocate the endpoint.
The endpoint type $\SessionEnd$ denotes an endpoint on which no
further input/output operation is possible and that can be deallocated.
An endpoint type
$\InternalChoice{\tmsg{\Tag_i}{\tbound{\tvar_i}{\TypeT_i}}{\TypeS_i}.\SessionType_i}_{i\in
  I}$ denotes an endpoint on which a process may send any message
$\Tag_i$ with $i\in I$. The message carries an argument of type
$\TypeS_i$ and the type variable $\tvar_i$ can be instantiated with
any subtype of the bound $\TypeT_i$ (subtyping will be defined
shortly).  Depending on the tag $\Tag_i$ of the message, the endpoint
can be used thereafter according to the endpoint type
$\SessionTypeT_i$.
In a dual manner, an endpoint type
$\ExternalChoice{\tmsg{\Tag_i}{\tbound{\tvar_i}{\TypeT_i}}{\TypeS_i}.\SessionType_i}_{i\in
  I}$ denotes and endpoint from which a process must be ready to
receive any message $\Tag_i$ with $i\in I$. Again, $\TypeS_i$ denotes
the type of the message's argument, while $\TypeT_i$ is the bound for
the type variable $\tvar_i$. Depending on the tag $\Tag_i$ of the
received message, the endpoint is to be used according to
$\SessionTypeT_i$.
Terms $\tvar$ and $\trec\tvar.\SessionTypeT$ can be used to specify
recursive behaviors, as usual.
Types are either endpoint types or the top type $\Top$, which is
supertype of any other (endpoint) type.

Here are some handy conventions regarding types and endpoint types:
\begin{itemize}
\item we sometimes use an infix notation for internal and external
  choices and write
  ${!}\tmsg{\Tag_1}{\tbound{\tvar_1}{\TypeT_1}}{\TypeS_1}.\SessionTypeT_1
  \oplus \cdots \oplus
  {!}\tmsg{\Tag_n}{\tbound{\tvar_n}{\TypeT_n}}{\TypeS_n}.\SessionTypeT_n$
  instead of
  $\InternalChoice{\tmsg{\Tag_i}{\tbound{\tvar_i}{\TypeT_i}}{\TypeS_i}.\SessionTypeT_i}_{i\in\{1,\dots,n\}}$
  and
  ${?}\tmsg{\Tag_1}{\tbound{\tvar_1}{\TypeT_1}}{\TypeS_1}.\SessionTypeT_1
  + \cdots +
  {?}\tmsg{\Tag_n}{\tbound{\tvar_n}{\TypeT_n}}{\TypeS_n}.\SessionTypeT_n$
  instead of
  $\ExternalChoice{\tmsg{\Tag_i}{\tbound{\tvar_i}{\TypeT_i}}{\TypeS_i}.\SessionTypeT_i}_{i\in\{1,\dots,n\}}$;

\item we omit the bound when it is $\Top$ and write, for example,
  ${!}\tmsg\Tag\tvar\TypeS.\SessionTypeT$ in place of
  ${!}\tmsg\Tag{\tbound\tvar\Top}\TypeS.\SessionTypeT$;

\item we omit the bound specification $\langle\cdot\rangle$
  altogether when useless (if the type variable occurs nowhere else)
  and write, for example, ${!}\Tag(\TypeS).\SessionTypeT$;

\item we omit the type of the argument when irrelevant.
\end{itemize}

We have standard notions of free and bound type variables for
(endpoint) types. The only binders are $\trec$ and the bound
constraints for messages. In particular, $\trec\tvar.\SessionTypeT$
binds $\tvar$ in $\SessionTypeT$ and
${\dagger}\tmsg\Tag{\tbound{\tvar}{\TypeT}}\TypeS.\SessionTypeT$ where
${\dagger} \in \{ {!}, {?} \}$ binds $\tvar$ in $\TypeS$ and in
$\SessionTypeT$ but not in $\TypeT$.
We adopt the Barendregt convention for type variables.
In what follows we will consider endpoint types modulo renaming of
bound variables and folding/unfolding of recursions, that is
$\trec\tvar.\SessionType =
\SessionType\subst{\trec\tvar.\SessionType}{\tvar}$ where
$\SessionType\subst{\trec\tvar.\SessionType}{\tvar}$ is the endpoint
type obtained from $\SessionType$ by replacing each free occurrence of
$\tvar$ with $\trec\tvar.\SessionType$.

The duality between inputs and outputs induces dual bounding
modalities.
In particular, a process using an endpoint with type
${!}\tmsg\Tag{\tbound{\tvar}{\TypeT}}\TypeS.\SessionTypeT$ may choose
a particular $\TypeT' \subt \TypeT$ to instantiate $\tvar$ and use
$\SessionTypeT\subst{\TypeT'}{\tvar}$ accordingly. In other words, the
variable $\tvar$ is \emph{universally quantified} over all the
subtypes of $\TypeT$.
Dually, a process using an endpoint with type
${?}\tmsg\Tag{\tbound{\tvar}{\TypeT}}\TypeS.\SessionTypeT$ does not
know the exact type $\TypeT' \subt \TypeT$ with which $\tvar$ is
instantiated, since this type is chosen by the sender. In other words,
the type variables of an external choice are \emph{existentially
  quantified} over all the subtypes of the corresponding bounds.

Not every term generated by the grammar in Table~\ref{tab:type_syntax}
makes sense.
Type variables bound by a recursion $\trec$ must be guarded by a
prefix (therefore a non-contractive endpoint type such as
$\trec\tvar.\tvar$ is forbidden) and type variables bound in a
constraint as in
${!}\tmsg\Tag{\tbound\tvar\TypeT}\TypeS.\SessionTypeT$ can only occur
in $\TypeS$ and within the prefixes of $\SessionTypeT$.
We formalize this last requirement as a well-formedness condition for
types denoted by a judgment $\innervars, \outervars \vdash \Type$
where $\innervars$ is a set of so-called \emph{inner variables} (those
that can occur only within prefixes) and $\outervars$ is a set of
so-called \emph{outer variables} (those that can occur everywhere):
\[
\begin{array}{c}
\inferrule{}{
  \innervars, \outervars \vdash \SessionEnd
}
\qquad
\inferrule{}{
  \innervars, \outervars \vdash \Top
}
\qquad
\inferrule{
  \tvar \in \outervars \setminus \innervars
}{
  \innervars, \outervars \vdash \tvar
}
\qquad
\inferrule{
  \innervars, \outervars \cup \{ \tvar \} \vdash \SessionType
}{
  \innervars, \outervars \vdash \trec\tvar.\SessionType
}
\\\\
\inferrule{
  \dagger\in\{{!},{?}\}
  \\
  \emptyset, \innervars \cup \outervars \vdash \TypeT_i
  ~{}^{(i\in I)}
  \\
  \emptyset, \innervars \cup \outervars \cup \{ \tvar_i \} \vdash \TypeS_i
  ~{}^{(i\in I)}
  \\
  \innervars \cup \{ \tvar_i \}, \outervars \vdash \SessionType_i
  ~{}^{(i\in I)}
}{
  \innervars, \outervars \vdash {\dagger}\{\tmsg{\Tag_i}{\tbound{\tvar_i}{\TypeT_i}}{\TypeS_i}.\SessionType_i\}_{i\in I}
}
\end{array}
\]

We say that $\Type$ is well formed if $\emptyset, \emptyset \vdash
\Type$ is derivable. Observe that well-formed (endpoint) types are
closed.
Well formedness restricts the expressiveness of types, in particular
types such as ${!}\tmsg\Tag{\tbound\tvar\TypeT}\TypeS.\tvar$ and
${?}\tmsg\Tag{\tbound\tvar\TypeT}\TypeS.\tvar$ are forbidden because
ill formed. We claim that ill-formed types have negligible practical
utility: a process that instantiates $\tvar$ with $\SessionTypeS \subt
\SessionTypeT$ in
${!}\tmsg\Tag{\tbound\tvar\SessionTypeT}\TypeS.\tvar$ precisely knows
its behavior after the output operation; dually, a process that
receives a message from an endpoint with type
${?}\tmsg\Tag{\tbound\tvar\SessionTypeT}\TypeS.\tvar$ cannot do any
better than assuming that $\tvar$ has been instantiated with
$\SessionTypeT$.

Duality expresses the fact that two processes accessing peer endpoints
interact without errors if they behave in complementary ways: if one
of the two processes sends a message, the other process waits for a
message; if one process waits for a message, the other process sends;
if one process has finished using an endpoint, the other process has
finished too.

\begin{definition}[duality]
  We say that $\drel$ is a \emph{duality relation} if $(\SessionTypeT,
  \SessionTypeS) \in {\drel}$ implies either
\begin{itemize}
\item $\SessionTypeT = \SessionTypeS = \SessionEnd$, or

\item $\SessionTypeT =
  \ExternalChoice{\tmsg{\Tag_i}{\tbound{\tvar_i}{\TypeT_i}}{\TypeS_i}.\SessionTypeT_i}_{i\in
    I}$ and $\SessionTypeS =
  \InternalChoice{\tmsg{\Tag_i}{\tbound{\tvar_i}{\TypeT_i}}{\TypeS_i}.\SessionTypeS_i}_{i\in
    I}$ and $(\SessionTypeT_i, \SessionTypeS_i) \in {\drel}$ for every
  $i\in I$,

\item $\SessionTypeT =
  \InternalChoice{\tmsg{\Tag_i}{\tbound{\tvar_i}{\TypeT_i}}{\TypeS_i}.\SessionTypeT_i}_{i\in
    I}$ and $\SessionTypeS =
  \ExternalChoice{\tmsg{\Tag_i}{\tbound{\tvar_i}{\TypeT_i}}{\TypeS_i}.\SessionTypeS_i}_{i\in
    I}$ and $(\SessionTypeT_i, \SessionTypeS_i) \in {\drel}$ for every
  $i\in I$.
\end{itemize}

We say that $\SessionTypeT$ and $\SessionTypeS$ are \emph{dual} if
$(\SessionTypeT, \SessionTypeS) \in {\drel}$ for some duality relation
$\drel$.
\end{definition}

It is easy to see that if $\SessionTypeT$ and $\SessionTypeS_1$ are
dual and $\SessionTypeT$ and $\SessionTypeS_2$ are dual, then
$\SessionTypeS_1 = \SessionTypeS_2$. In other words, the duality
relation induces a function $\co{\,\cdot\,}$ such that $\SessionTypeT$
and $\co\SessionTypeT$ are dual for every $\SessionTypeT$.

An important property of well-formed endpoint types is that duality
does not affect their inner variables. Therefore, duality and the
instantiation of inner variables commute, in the following sense:

\begin{proposition}
  \label{prop:subst_dual}
  Let $\{ \tvar \}, \emptyset \vdash \SessionTypeT$. Then
  $\co{\SessionTypeT\subst{\TypeS}{\tvar}} =
  \co{\SessionTypeT}\subst{\TypeS}{\tvar}$.
\end{proposition}

We now define subtyping. Because of bound constraints on type
variables, subtyping is relative to an environment $\BoundContext =
\tbound{\tvar_1}{\TypeT_1},\dots,\tbound{\tvar_n}{\TypeT_n}$ which is
an ordered sequence of constraints such that each $\tvar_i$ may only
occur in the $\TypeT_j$'s with $j > i$.
We write $\dom(\BoundContext)$ for the domain of $\BoundContext$ and
$\BoundContext(\tvar_i)$ to denote the bound $\TypeT_i$ associated
with the rightmost occurrence of $\tvar_i\in\dom(\BoundContext)$; we
use $\EmptyBoundContext$ to denote the empty bound environment.
Subtyping is fairly standard, therefore we provide only a coinductive
characterization (an equivalent deduction system restricted to finite
endpoint types can be found in~\cite{Gay08}).

\begin{definition}[subtyping]
\label{def:subt}
We say that $\srel$ is a \emph{coinductive subtyping} if
$(\BoundContext, \TypeT, \TypeS) \in {\srel}$ implies either:
\begin{enumerate}
\item $\TypeT = \TypeS$, or

\item $\TypeS = \Top$, or

\item $\TypeT = \tvar \in \dom(\BoundContext)$ and $(\BoundContext,
  \BoundContext(\tvar), \TypeS) \in {\srel}$, or

\item $\TypeT =
  \ExternalChoice{\tmsg{\Tag_i}{\tbound{\tvar_i}{\TypeT_i}}{\TypeS_i}.\SessionTypeT_i}_{i\in
    I}$ and $\TypeS =
  \ExternalChoice{\tmsg{\Tag_i}{\tbound{\tvar_i}{\TypeT_i}}{\TypeS_i'}.\SessionTypeS_i}_{i\in
    J}$ with $I\subseteq J$ and $((\BoundContext,
  \tbound{\tvar_i}{\TypeT_i}), \TypeS_i, \TypeS_i') \in {\srel}$ and
  $((\BoundContext, \tbound{\tvar_i}{\TypeT_i}), \SessionTypeT_i,
  \SessionTypeS_i) \in {\srel}$ for every $i\in I$, or

\item $\TypeT =
  \InternalChoice{\tmsg{\Tag_i}{\tbound{\tvar_i}{\TypeT_i}}{\TypeS_i}.\SessionTypeT_i}_{i\in
    I}$ and $\TypeS =
  \InternalChoice{\tmsg{\Tag_i}{\tbound{\tvar_i}{\TypeT_i}}{\TypeS_i'}.\SessionTypeS_i}_{i\in
    J}$ with $J\subseteq I$ and $((\BoundContext,
  \tbound{\tvar_i}{\TypeT_i}), \TypeS_i', \TypeS_i) \in {\srel}$ and
  $((\BoundContext, \tbound{\tvar_i}{\TypeT_i}), \SessionTypeT_i,
  \SessionTypeS_i) \in {\srel}$ for every $i\in I$.
\end{enumerate}

We write $\BoundContext \vdash \TypeT \subt \TypeS$ if
$(\BoundContext, \TypeT, \TypeS) \in {\srel}$ for some coinductive
subtyping $\srel$ and $\TypeT \subt \TypeS$ if $\EmptyBoundContext
\vdash \TypeT \subt \TypeS$.
\end{definition}

Item~(1) states that subtyping is reflexive;
item~(2) states that $\Top$ is indeed the top type;
items~(4) and~(5) are the usual covariant and contravariant rules for
inputs and outputs respectively. Observe that subtyping is always
covariant with respect to the continuations and that we require the
bounds on type variables of related endpoint types be the same. This
corresponds to the so-called ``Kernel'' variant of bounded
polymorphism as opposed to the ``Full'' one. We adopt the Kernel
variant for simplicity, since it is orthogonal to the rest of the
theory. Also, the Full variant is known to be
undecidable~\cite{Gay08}.
Finally, item~(3) allows one to deduce $\BoundContext \vdash \tvar
\subt \TypeS$ if $\BoundContext \vdash \BoundContext(\tvar) \subt
\TypeS$ holds.

The reader may easily verify that subtyping is transitive (it is
enough to show that $\{ (\BoundContext, \TypeT_1, \Type_2) \mid
\exists \TypeS: \BoundContext \vdash \TypeT_1 \subt \TypeS \land
\BoundContext \vdash \TypeS \subt \TypeT_2 \}$ is a coinductive
subtyping).
The following property is standard and shows that duality is
contravariant with respect to subtyping.

\begin{proposition}
\label{prop:subt_dual}
$\SessionTypeT \subt \SessionTypeS$ if and only if $\co\SessionTypeS
\subt \co\SessionTypeT$.
\end{proposition}

Well-formedness of endpoint types is essential for
Proposition~\ref{prop:subt_dual} to hold in our setting.  Consider,
for example, the endpoint types $\SessionTypeT =
{!}\tmsg{\Tag}{\tbound{\tvar}{{?}\Tag().\SessionEnd}}{}.\tvar$ and
$\SessionTypeS =
{!}\tmsg{\Tag}{\tbound{\tvar}{{?}\Tag().\SessionEnd}}{}.{?}\Tag().\SessionEnd$
where $\SessionTypeT$ is ill formed.
Then $\SessionTypeT \subt \SessionTypeS$ would hold but
${?}\tmsg{\Tag}{\tbound{\tvar}{{?}\Tag().\SessionEnd}}{}.{!}\Tag().\SessionEnd
= \co\SessionTypeS \subt \co\SessionTypeT =
{?}\tmsg{\Tag}{\tbound{\tvar}{{?}\Tag().\SessionEnd}}{}.\tvar$ would
not because $\tbound{\tvar}{{?}\Tag().\SessionEnd} \vdash
{!}\Tag().\SessionEnd \not\subt \tvar$.
An alternative theory where well-formedness is not necessary for
proving Proposition~\ref{prop:subt_dual} is given in~\cite{Gay08} and
consists in distinguishing dualized type variables $\co\tvar$ from
type variables and by having type constraints of the form $\TypeT_1
\subt \tvar \subt \TypeT_2$ with both lower and upper bounds, so that
the bounds of the corresponding dualized type variable are known and
given by $\co{\TypeT_2} \subt \co\tvar \subt \co{\TypeT_1}$.

\paragraph{Typing the heap.}
The heap plays a primary role in our setting because inter-process
communication utterly relies on heap-allocated structures; also, most
properties of well-behaved processes are direct consequences of
related properties of the heap.
Therefore, just as we will check well typedness of a process
$\Process$ with respect to some environment that associates the
pointers occurring in $\Process$ with the corresponding types, we will
also need to check that the heap is consistent with respect to the
same environment. This leads to a notion of well-typed heap that we
develop in this section.
The mere fact that we have this notion does not mean that we need to
type-check the heap at runtime. Well typedness of the heap will be a
consequence of well typedness of processes, and the empty heap will be
trivially well typed.
We will express well-typedness of a heap $\Memory$ with respect to a
pair $\Context_0;\Context$ of environments where $\Context$ represents
the type of the \emph{roots} of $\Memory$ (the pointers that are not
referenced by any other structure allocated in the heap) and
$\Context_0$ describes the type of the pointers to allocated
structures that are directly or indirectly reachable from one of the
roots of the heap.

Among the properties that a well-typed heap must enjoy is the
complementarity between the endpoint types associated with peer
endpoints. This notion of complementarity does not coincide with
duality because of the communication model that we have adopted, which
is asynchronous. Since messages can accumulate in the queue of an
endpoint before they are received, the type of the endpoint as
perceived by the process using it and the actual type of the endpoint
as assumed by the process using its peer can be misaligned.
On the one hand, we want to enforce the invariant that the endpoint
types of peer endpoints are (and remain) dual, modulo the subtyping
relation. On the other hand, this can only happen when the two
endpoints have empty queues. In general, we need to compute the actual
endpoint type of an endpoint taking into account its type as perceived
by the process using it \emph{and} the messages in its queue.
This is accomplished by the $\tail({\cdot}, {\cdot})$ function below,
which takes an endpoint type $\SessionTypeT$ and a sequence
$\Tag_1(\TypeS_1)\cdots\Tag_n(\TypeS_n)$ of specifications
corresponding the messages enqueued into the endpoint with type
$\SessionTypeT$ and computes the residual endpoint type that assumes
that all those messages have been received:
\[
\inferrule{}{
  \tail(\SessionType, \varepsilon) = \SessionType
}
\qquad
\inferrule{
  k\in I
  \\
  \TypeT \subt \TypeT_k
  \\
  \TypeS \subt \TypeS_k\subst{\TypeT}{\tvar_k}
  \\
  \tail(\SessionTypeT_k\subst{\TypeT}{\tvar_k}, \Tag_1(\TypeS_1')\cdots\Tag_n(\TypeS_n')) = \SessionTypeS
}{
  \tail(\ExternalChoice{\tmsg{\Tag_i}{\tbound{\tvar_i}{\TypeT_i}}{\TypeS_i}.\SessionTypeT_i}_{i\in I}, \Tag_k(\TypeS)\Tag_1(\TypeS_1')\cdots\Tag_n(\TypeS_n'))
  = 
  \SessionTypeS
}
\]

From a technical point of view $\tail$ is a relation, since there can
be several possible choices for instantiating the type variables in
the endpoint type being processed. For example, we have
$\tail({?}\tmsg\Tag{\tbound{\tvar}{\TypeT}}\tvar.{?}\Tag(\tvar).\SessionEnd,
\Tag(\TypeS)) = {?}\Tag(\TypeT').\SessionEnd$ for every $\TypeS \subt
\TypeT' \subt \TypeT$. Nonetheless we will sometimes use $\tail$ as a
function and write $\tail(\SessionTypeT,
\Tag_1(\TypeS_1)\cdots\Tag_m(\TypeS_n))$ in place of some
$\SessionTypeS$ such that $\tail(\SessionTypeT,
\Tag_1(\TypeS_1)\cdots\Tag_m(\TypeS_n)) = \SessionTypeS$. For
instance, the notation $\tail(\SessionTypeT,
\Tag_1(\TypeS_1)\cdots\Tag_m(\TypeS_n)) \subt \SessionTypeS'$ means
that $\SessionTypeS \subt \SessionTypeS'$ for some $\SessionTypeS$
such that $\tail(\SessionTypeT,
\Tag_1(\TypeS_1)\cdots\Tag_m(\TypeS_n)) = \SessionTypeS$.

We now have all the notions to express the well-typedness of a heap
$\Memory$ with respect to a pair $\Context_0; \Context$ of type
environments.

\begin{definition}[well-typed heap]
\label{def:wth}
We write $\Context_0;\Context \vdash \Memory$ if:
\begin{enumerate}
\item for every $\PointerA \mapsto [\PointerB, \Queue] \in \Memory$ we
  have $\PointerB \mapsto [\PointerA, \Queue'] \in \Memory$ and either
  $\Queue = \EmptyQueue$ or $\Queue' = \EmptyQueue$;

\item for every $\PointerA \mapsto [\PointerB,
  \msg{\Tag_1}{\PointerC_1}::\cdots::\msg{\Tag_n}{\PointerC_n}] \in
  \Memory$ and $\PointerB \mapsto [\PointerA, \EmptyQueue]$ we have
  $\co{\tail(\SessionTypeT, \Tag_1(\TypeS_1)\cdots\Tag_m(\TypeS_n))}
  \subt \SessionTypeS$ where $\PointerA : \SessionTypeT \in \Context$
  and $\PointerB : \SessionTypeS \in \Context$ and $\PointerC_i :
  \TypeS_i \in \Context$ for $i\in\{1,\dots, n\}$;

\item $\dom(\Memory) = \dom(\Context_0,\Context) =
  \reachable{\dom(\Context)}\Memory$;

\item $\reachable{\{\PointerA\}}\Memory \cap
  \reachable{\{\PointerB\}}\Memory = \emptyset$ for every
  $\PointerA,\PointerB \in \dom(\Context)$ with
  $\PointerA\ne\PointerB$.
\end{enumerate}
\end{definition}

Condition~(1) requires that in a well-typed heap every endpoint comes
along with its peer and that at least one of the queues of peer
endpoints be empty. This invariant is ensured by duality, since a
well-typed process does not send messages on an endpoint until it has
read all the pending messages from the corresponding queue.
Condition~(2) requires that the endpoint types of peer endpoints are
dual, modulo subtyping. More precisely, for every endpoint $\PointerA$
whose peer $\PointerB$ has an empty queue there exists a residual
$\tail(\SessionTypeT, \Tag_1(\TypeS_1)\cdots\Tag_n(\TypeS_n))$ of its
type $\SessionTypeT$ whose dual is a subtype of the peer's type
$\SessionTypeS$.
Condition~(3) states that the type environment $\Context_0,\Context$
must specify a type for all of the allocated objects in the heap and,
in addition, every object (located at) $\PointerA$ in the heap must be
reachable from a root $\PointerB \in \dom(\Context)$. Since the roots
will be distributed linearly to the processes of the system, this
guarantees the absence of leaks, namely of allocated objects which are
no longer reachable.
Finally, condition~(4) requires the uniqueness of the root for every
allocated object. This guarantees process isolation, namely the fact
that every allocated object belongs to one and only one process.

\paragraph{Typing processes.}
We want to define a type system such that well-typed processes are
well behaved and, in particular, such that well-typed processes do not
leak memory. As we have anticipated in the introduction, the critical
situation that we must avoid is the possibility that an endpoint is
enqueued into its own queue, since this would cause the creation of a
circular structure not owned by any process.

The intuition behind our solution is to use some property of types to
detect --- and avoid --- the configurations in which there exists some
potential to create such circular structures.
This property, which we dub \emph{weight} of a type, gives an upper
bound to the length of chains of pointers linking endpoints.

\begin{definition}[weight]
  We say that $\wrel$ is a \emph{coinductive weight bound} if
  $(\BoundContext, \Type, n) \in \wrel$ implies either:
\begin{itemize}
\item $\Type = \SessionEnd$, or

\item $\Type =
  \InternalChoice{\tmsg{\Tag_i}{\tbound{\tvar_i}{\TypeT_i}}{\TypeS_i}.\SessionType_i}_{i\in
    I}$, or

\item $\Type = \tvar \in \dom(\BoundContext)$ and $(\BoundContext,
  \BoundContext(\tvar), n) \in \wrel$, or

\item $\Type =
  \ExternalChoice{\tmsg{\Tag_i}{\tbound{\tvar_i}{\TypeT_i}}{\TypeS_i}.\SessionType_i}_{i\in
    I}$ and $n > 0$ and $((\BoundContext, \tbound{\tvar_i}{\TypeT_i}),
  \TypeS_i, n-1) \in \wrel$ and $((\BoundContext,
  \tbound{\tvar_i}{\TypeT_i}), \SessionType_i, n) \in \wrel$ for every
  $i\in I$.
\end{itemize}

We write $\wbb\BoundContext\Type{n}$ if $(\BoundContext,
\Type, n) \in {\wrel}$ for some coinductive weight bound $\wrel$.
The \emph{weight} of a type $\Type$ with respect to $\BoundContext$,
denoted by $\xweight\BoundContext\Type$, is defined by
$\xweight\BoundContext\Type = \min\{n\in\natset \mid
\wbb\BoundContext\Type{n}\}$ where we let $\min\emptyset = \infty$.
We omit the environment $\BoundContext$ when it is empty and simply
write $\weight\Type$ instead of $\xweight\EmptyBoundContext\Type$.
When comparing weights, we extend the usual total orders $<$ and
$\leq$ over natural numbers so that $n < \infty$ for every
$n\in\natset$ and $\infty \leq \infty$.
\end{definition}

Like other relations involving types, the relation
$\wbb\BoundContext\Type{n}$ is parametric on an environment
$\BoundContext$ specifying the upper bound of type variables that may
occur in $\Type$ and expresses the fact that $n$ is an upper bound for
the weight of $\Type$. The weight of $\Type$ is then defined as the
least of its upper bounds, or $\infty$ if there is no such upper
bound.
A few weights are straightforward to compute, for example we have
$\weight\SessionEnd =
\weight{\InternalChoice{\tmsg{\Tag_i}{\tbound{\tvar_i}{\TypeT_i}}{\TypeS_i}.\SessionType_i}_{i\in
    I}} = 0$ and $\weight\Top = \infty$. Endpoints with type
$\SessionEnd$ and those in a send state have a null weight because
their corresponding queues are empty and therefore the chains of
pointers originating from them has zero length. In the case of $\Top$,
it does not have a finite weight since $\Top$ is the type of
\emph{any} endpoint, and in particular of any endpoint with an
arbitrary weight. Only endpoints in a receive state do have a strictly
positive weight. For instance we have
$\weight{{?}\Tag(\SessionEnd).\SessionEnd} = 1$ and
$\weight{{?}\Tag({?}\Tag(\SessionEnd).\SessionEnd).\SessionEnd} = 2$,
while $\xweight\BoundContext{\trec\tvar.{?}\Tag(\tvar).\SessionTypeT}
= \xweight\BoundContext{{?}\Tag(\Top).\SessionTypeT} =
\infty$.
The weight of type variables occurring in a constraint
$\tbound\tvar\Type$ is given by the weight of $\Type$. In particular,
$\xweight{\tbound\tvar\Type}\tvar = \weight\Type$ and
$\weight{{?}\tmsg\Tag{\tbound\tvar\Type}\tvar.\SessionEnd} = 1 +
\weight\Type$ (this latter equality holds if $\weight\Type < \infty$).
In a sense, the (type) bound $\Type$ for $\tvar$ determines also a
(weight) bound $\weight\Type$ for $\tvar$. Since the actual type with
which $\tvar$ will be instantiated is not known, this approximation
works well only if the relation between the weights is coherent with
subtyping. This fundamental property does indeed hold, as stated in
the following proposition.

\begin{proposition}
\label{prop:subt_weight}
$\TypeT \subt \TypeS$ implies $\weight\TypeT \leq \weight\TypeS$.
\end{proposition}

\begin{table}
\caption{\label{tab:typing_processes} Typing rules for processes.}
\framebox[\textwidth]{
\begin{math}
\displaystyle
\begin{array}{@{}c@{}}
  \inferrule[\rulename{T-Idle}]{}{
    \RecContext; \BoundContext; \emptyset \vdash \idle
  }
  \qquad
  \inferrule[\rulename{T-Close}]{}{
    \RecContext; \BoundContext; \Name : \SessionEnd \vdash \closeChannel(\Name)
  }
  \qquad
  \inferrule[\rulename{T-Rec}]{
    \RecContext, \{ \RecVar \mapsto (\BoundContext; \Context) \};
    \BoundContext;
    \Context \vdash \Process
    \\
    \dom(\Context) = \fn(\Process) 
  }{
    \RecContext; \BoundContext; \Context \vdash \rec\RecVar.\Process
  }
  \\\\
  \inferrule[\rulename{T-Open}]{
    \RecContext; \BoundContext; \Context, \PointerA : \SessionTypeT, \PointerB : \co\SessionTypeT
    \vdash
    \Process
  }{
    \RecContext; \BoundContext; \Context
    \vdash
    \openChannel(\PointerA, \PointerB).
    \Process
  }
  \qquad
  \inferrule[\rulename{T-Send}]{
    \BoundContext \vdash \TypeT' \subt \TypeT
    \\
    \xweight\BoundContext{\TypeS\subst{\TypeT'}{\tvar}} < \infty
    \\
    \RecContext;
    \BoundContext;
    \Context, \NameA : \SessionTypeS\subst{\TypeT'}{\tvar}
    \vdash
    \Process
  }{
    \RecContext;
    \BoundContext;
    \Context, \NameA : {!}\tmsg{\Tag}{\tbound{\tvar}{\TypeT}}{\TypeS}.\SessionTypeS, \NameB : \TypeS\subst{\TypeT'}{\tvar}
    \vdash
    \send\NameA{\Tag}{\NameB}.\Process
  }
  \\\\
  \inferrule[\rulename{T-Var}]{}{
    \RecContext, \{ \RecVar \mapsto (\BoundContext; \Context) \};
    \BoundContext;
    \Context \vdash \RecVar
  }
  \qquad
  \inferrule[\rulename{T-Receive}]{
    \RecContext;
    \BoundContext, \tbound{\tvar_i}{\TypeT_i};
    \Context, \NameA : \SessionType_i, \Var_i : \TypeS_i
    \vdash
    \Process_i
    ~{}^{(i\in I)}
  }{
    \textstyle
    \RecContext;
    \BoundContext;
    \Context,
    \NameA : \ExternalChoice{\tmsg{\Tag_i}{\tbound{\tvar_i}{\TypeT_i}}{\TypeS_i}.\SessionType_i}_{i\in I}
    \vdash
    \sum_{i\in I} \NameA?\Tag_i(\Var_i).\Process_i
  }
  \\\\
  \inferrule[\rulename{T-Choice}]{
    \RecContext; \BoundContext; \Context \vdash \ProcessP
    \\
    \RecContext; \BoundContext; \Context \vdash \ProcessQ
  }{
    \RecContext; \BoundContext; \Context \vdash \ProcessP \choice \ProcessQ
  }
  \qquad
  \inferrule[\rulename{T-Par}]{
    \RecContext; \BoundContext; \Context_1 \vdash \ProcessP
    \\
    \RecContext; \BoundContext; \Context_2 \vdash \ProcessQ
  }{
    \RecContext; \BoundContext; \Context_1, \Context_2 \vdash \ProcessP \parop \ProcessQ
  }
  \qquad
  \inferrule[\rulename{T-Sub}]{
    \RecContext; \BoundContext; \Context, \Name : \TypeS \vdash \Process
    \\
    \BoundContext \vdash \TypeT \subt \TypeS
  }{
    \RecContext; \BoundContext; \Context, \Name : \TypeT \vdash \Process
  }
\end{array}
\end{math}
}
\end{table}

The typing rules for processes are inductively defined in
Table~\ref{tab:typing_processes}. Judgments have the form
$\RecContext; \BoundContext; \Context \vdash \Process$ and state that
process $\Process$ is well typed under the specified environments.
The additional environment $\RecContext$ is a map from process
variables to pairs $(\BoundContext; \Context)$ and is used for typing
recursive processes. It plays a role in two rules only,
\rulename{T-Var} and~\rulename{T-Rec}, which are standard except for
the unusual premise $\dom(\Context) = \fn(\Process)$ in
rule~\rulename{T-Rec} that enforces a weak form of contractivity on
recursive processes. It states that $\rec\RecVar.\Process$ is well
typed under $\Context$ only if $\Process$ actually uses the names in
$\dom(\Context)$. Normally, divergent processes such as
$\rec\RecVar.\RecVar$ can be typed in every environment.  If this were
the case, the process $\openChannel(\PointerA,
\PointerB).\rec\RecVar.\RecVar$, which leaks $\PointerA$ and
$\PointerB$, would be well typed.
The idle process is well typed in the empty environment
$\EmptyContext$. Since we will impose a correspondence between the
free names of a process and the roots of the heap, this rule states
that the terminated process has no leaks.
Rule~\rulename{T-Close} states that a process $\closeChannel(\Name)$
is well typed provided that $\Name$ is the only name owned by the
process and that it corresponds to an endpoint with type
$\SessionEnd$, on which no further interaction is possible.
Rule~\rulename{T-Open} types the creation of a new channel, which is
visible in the continuation process as two peer endpoints typed by
dual endpoint types.
Rules~\rulename{T-Choice} and~\rulename{T-Par} are standard. In the
latter, the type environment is split into disjoint environments to
type the processes being composed. Together with heap well-typedness,
this ensures process isolation.
Rule~\rulename{T-Send} states that a process
$\NameA!\msg{\Tag}{\NameB}.\Process$ is well typed if $\NameA$ is
associated with an endpoint type
${!}\tmsg{\Tag}{\tbound{\tvar}{\TypeT}}{\TypeS}.\SessionTypeS$ that
permits the output of $\Tag$-tagged messages. The rule guesses the
type parameter $\TypeT'$ with which the type variable $\tvar$ is
instantiated (an explicitly typed process might explicitly provide
$\TypeT'$). The type of the argument $\NameB$ must match the expected
type in the endpoint type where $\tvar$ has been instantiated with
$\TypeT'$ and the continuation $\Process$ must be well typed in a
context where the message argument has disappeared and the endpoint
$\NameA$ is typed according to a properly instantiated
$\SessionTypeS$. This means that $\Process$ can rely on the knowledge
of $\TypeT'$, namely $\tvar$ is universally quantified over all the
subtypes of $\TypeT$.
The condition $\xweight\BoundContext{\TypeS\subst{\TypeT'}{\tvar}} <
\infty$ imposes that $\NameB$'s type must have a finite weight. Since
the peer of $\NameA$ must be able to accept a message with an argument
of type $\TypeS\subst{\TypeT'}{\tvar}$, its weight will be strictly
larger than that of $\TypeS\subst{\TypeT'}{\tvar}$, or it will be
infinite. In both cases, we are sure that the argument $\NameB$ being
sent is not the peer of $\NameA$.
Rule~\rulename{T-Receive} deals with inputs: a process waiting for a
message from an endpoint $\NameA :
\ExternalChoice{\tmsg{\Tag_i}{\tbound{\tvar_i}{\TypeT_i}}{\TypeS_i}.\SessionType_i}_{i\in
  I}$ is well typed if it can deal with any $\Tag_i$-tagged
message. The continuation process may use the endpoint $\NameA$
according to the endpoint type $\SessionType_i$ and can access the
message argument $\Var_i$. The environment $\Context$ is enriched with
the assumption $\tbound{\tvar_i}{\TypeT_i}$ denoting the fact that
$\Process_i$ does not know the exact type with which $\tvar_i$ has
been instantiated, but only its upper bound, namely $\tvar_i$ is
existentially quantified over all the subtypes of $\TypeT_i$.
Finally, rule~\rulename{T-Sub} is a subsumption rule for assumptions:
if a process $\Process$ is well typed with respect to a context
$\Context, \Name : \TypeS$, it remains well typed if the type
associated with $\Name$ is more precise than (is a subtype of)
$\TypeS$.

Systems $\system\Memory\Process$ are well typed if so are their
components:

\begin{definition}[well-typed system]
  We write $\Context_0;\Context \vdash (\Memory;\Process)$ if
  $\Context_0;\Context \vdash \Memory$ and $\Context \vdash \Process$.
\end{definition}

We conclude with two standard results about our framework:
well-typedness is preserved by reduction, and well-typed process are
well behaved.  Subject reduction is slightly non-standard, in the
sense that types in the environment may change as the process
reduces. This is common in session type theories, since session types
are behavioral types.

\begin{theorem}[subject reduction]
\label{thm:sr}
Let $\Context_0; \Context \vdash (\Memory; \Process)$ and $(\Memory;
\Process) \red{} (\Memory'; \Process')$. Then $\Context_0'; \Context'
\vdash (\Memory'; \Process')$ for some $\Context_0'$ and $\Context'$.
\end{theorem}

\begin{theorem}[safety]
\label{thm:safety}
Let $\vdash \Process$. Then $\Process$ is well behaved.
\end{theorem}

\begin{example}
\label{ex:cell_types}
Consider the endpoint type
\[
  \mathtt{CellT} =
  \trec\tvarA.
  ({!}\tmsg{\mathtt{set}}{\tvarB}{\tvarB}.{?}\mathtt{get}(\tvarB).\tvarA
  \oplus
  {!}\mathtt{free}().\SessionEnd)
\]
modeling the interface of a linear mutable cell. Then it is easy to
verify that
\[
\PointerC : \co{\mathtt{CellT}} \vdash \mathtt{CELL}(\PointerC)
\]
is derivable, where $\mathtt{CELL}$ is the process presented in
Example~\ref{ex:cell}. Therefore, $\mathtt{CELL}$ is a correct
implementation of a linear mutable cell.

Since $\mathtt{CellT}$ begins with an internal choice we have
$\weight{\mathtt{CellT}} = 0$. This means that it is always safe to
send an empty cell as the argument of a message since the second
premise of rule~\rulename{T-Send} will always be satisfied. On the
contrary, we have
$\xweight{\tbound\tvarB\Top}{{?}\mathtt{get}(\tvarB).\mathtt{CellT}} =
\infty$, therefore it seems like initialized cells can never be sent
as messages. However, if sender and initializer are the same process,
there might be just enough information to deduce that the process is
safe. For example, the judgment
\[
  \PointerA : \TypeT,
  \PointerB : {!}\mathtt{send}({?}\mathtt{get}(\TypeT).\mathtt{CellT}).\SessionEnd,
  \PointerC : \mathtt{CellT}
  \vdash
  \PointerC!\mathtt{set}(a).
  \PointerB!\mathtt{send}(\PointerC).
  \closeChannel(\PointerB)
\]
is derivable if $\weight\TypeT < \infty$. In this case, the
sub-process $\PointerB!\mathtt{send}(\PointerC).
\closeChannel(\PointerB)$ is type checked in an environment where the
(residual) endpoint type of $\PointerC$ has been instantiated to
${?}\mathtt{get}(\TypeT).\mathtt{CellT}$ whose weight is
$\weight{{?}\mathtt{get}(\TypeT).\mathtt{CellT}} = \weight\TypeT + 1 <
\infty$.
\eoe
\end{example}

\begin{example}
  Suppose we want to implement a forwarder process that receives two
  endpoints with dual types and forwards the stream of messages coming
  from the first endpoint to the second one. We might implement the
  process thus:
\[
  \mathtt{FWD}(\PointerA) =
    \PointerA?\mathtt{src}(x).
    \PointerA?\mathtt{dest}(y).
    \rec\RecVar.
    (x?\Tag(z).y!\Tag(z).\RecVar
    +x?\mathtt{eos}().y!\mathtt{eos}().(
    \closeChannel(x) \parop \closeChannel(y) \parop \closeChannel(\PointerA)))
\]

However, the judgment
\[
  \PointerA : {?}\tmsg{\mathtt{src}}{\tvarA}{\co{\mathtt{Stream}}}.
              {?}\mathtt{dest}({\mathtt{Stream}}).
              \SessionEnd
  \vdash
  \mathtt{FWD}(\PointerA)
\]
where $\mathtt{Stream} = \trec\tvarB.({!}\Tag(\tvarA).\tvarB \oplus
{!}\mathtt{eos}().\SessionEnd)$ is not derivable because there is no
upper bound to the weight of $\tvarA$ and $\mathtt{FWD}(\PointerA)$
attempts at sending $z$ where $z : \tvarA$.
Had $\mathtt{FWD}(\PointerA)$ been typable, it would be possible to
create a leak with the process
\[
  \openChannel(\PointerA, \PointerB).
  (\mathtt{FWD}(\PointerA) \parop
   \openChannel(\PointerC, \PointerD).
   \openChannel(\PointerE, \PointerF).
   \PointerB!\mathtt{src}(\PointerD).
   \PointerB!\mathtt{dest}(\PointerE).
   \PointerC!\Tag(\PointerF).
   \PointerC!\mathtt{eos}().
   (\closeChannel(\PointerB) \parop \closeChannel(\PointerC))
\]
which has the effect to enqueue $\PointerF$ into its own queue.  The
process $\mathtt{FWD}$ becomes typable as soon as $\tvarA$ in
$\PointerA$'s type is given a bound with a finite weight.
\eoe
\end{example}


\section{Related work}
\label{sec:related}

Copyless message passing is one of the key features adopted by the
Singularity OS~\cite{SingularityOverview05} to compensate the overhead
of communication-based interactions between isolated
processes. Communication safety and deadlock freedom can be ensured by
checking processes against \emph{channel contracts} that are
\emph{deterministic}, \emph{autonomous}, and
\emph{synchronizing}~\cite{StengelBultan09}.  As argued
in~\cite{Fahndrich06} and shown in~\cite{BonoMessaPadovani11}, the
first two conditions make it possible to split contracts into pairs of
dual endpoint types, and to implement the static analysis along the
lines of well known session type
theories~\cite{Honda93,HondaVasconcelosKubo98}.
In~\cite{BonoMessaPadovani11} it was also observed that these
conditions are insufficient for preventing memory leaks and it was
shown how to address the issue by imposing a ``finite-weight''
restriction to endpoint types. In the present paper, we further
generalize our solution by allowing infinite-weight endpoint types in
general, although only endpoints with finite-weight can actually be
sent as messages (see the extra premise of rule~\rulename{T-Send}).
As a matter of fact, \cite{Fahndrich06} already noted that the
implementation of ownership transfer posed some consistency issues if
endpoints not in a \emph{send-state} were allowed to be sent as
messages, but no relation with memory leaks was observed.  Since our
``finite-weight'' condition is a generalization of the
\emph{send-state} condition (\emph{send-state} endpoints always have
null weight), our work provides a formal proof that the
\emph{send-state} condition is sufficient also for avoiding memory
leaks.
Other works~\cite{Fahndrich06,GayVasconcelos10} introduce apparently
similar, ``finite-weight'' restrictions on session types to make sure
that message queues of the corresponding channels are bounded. Our
weights are unrelated to the size of queues and concern the length of
chains of pointers involving queues.

Polymorphic contracts and endpoint types of the Singularity
OS~\cite{SDN5} have never been formalized before nor do they appear to
be supported by the Singularity RDK.
Our polymorphic endpoint types are closely related to the session
types in \cite{Gay08}, which was the first work to introduce bounded
polymorphism for session types. There are a few technical differences
between our polymorphic endpoint types and the session types
in~\cite{Gay08}: we unify external and internal choices respectively
with input and output operations, so as to model Singularity contracts
more closely; we admit recursive endpoint types, while the support for
recursion was only informally sketched in~\cite{Gay08}. This led us to
define subtyping coinductively, rather than by means of an inductive
deduction system. Finally, we dropped lower type bounds and preferred
to work with a restricted language of well-formed endpoint types which
we claim to be appropriate in practice.
Another work where session types are enriched with bounded
polymorphism is~\cite{DDGY07}, but in that case polymorphism is
restricted to data, while in~\cite{Gay08} and in the present paper
type variables range over behaviors as well.
Interestingly, all the critical endpoint types
in~\cite{BonoMessaPadovani11} that violate the ``finite-weight''
restriction are recursive. This is no longer the case when (bounded)
polymorphism is added and in fact there are finite endpoint types that
can cause memory leaks if sent as messages.

A radically different approach for the static analysis of Singularity
processes is given
by~\cite{VillardLozesCalcagno09,VillardLozesCalcagno10}, where the
authors develop a proof system based on a variant of \emph{separation
  logic}~\cite{OHearnReynoldsYang01}.  
However, leaks in~\cite{VillardLozesCalcagno09} manifest themselves
only when both endpoints of any channel have been closed. In
particular, it is possible to prove that the code fragment
\eqref{micidiale} is correct, although it does indeed leak some
memory.
This problem has been subsequently recognized and solved
in~\cite{Villard11}. \REVISION{Roughly, the solution consists in
  forbidding the output of a message unless it is possible to prove
  (in the logic) that the queue that is going to host the message is
  reachable from the content of the message itself. In principle this
  condition is optimal, in the sense that it should permit every safe
  output. However, it relies on the knowledge of the identity of
  endpoints, that is a very precise information that is not always
  available. For this reason, \cite{Villard11} also proposes an
  approximation of this condition, consisting in tagging endpoints of
  a channel with distinct \emph{roles} (basically, what are called
  \emph{importing} and \emph{exporting} ends in Singularity). Then, an
  endpoint can be safely sent as a message only if its role matches
  the one of the endpoint on which it is sent. This solution is
  incomparable to the one we advocate -- restricting the output to
  endpoints with finite-weight type -- suggesting that it may be
  possible to work out a combination of the two.}

\REVISION{There exist a few works on session
  types~\cite{BCDDDY08,CastagnaDezaniGiachinoPadovani09} that
  guarantee a global progress property for well-typed systems where
  the basic idea is to impose an order on channels to prevent circular
  dependencies that could lead to a deadlock.
  Not surprisingly, the critical processes (such as~\eqref{micidiale})
  that we rule out thanks to the finite-weight restriction on the type
  of messages are ill typed in these works.
  It turns out that a faithful encoding of~\eqref{micidiale} into the
  models proposed in these works is impossible, because the
  $\openChannel({\cdot},{\cdot})$ primitive we adopt creates
  \emph{both} endpoints of a channel within the same process, while
  the session initiation primitives
  in~\cite{BCDDDY08,CastagnaDezaniGiachinoPadovani09} associate the
  fresh endpoints of a newly opened session to different processes
  running in parallel. This invariant -- that the same process cannot
  own more than one endpoint of the same channel -- is preserved in
  well-typed processes because of a severe restriction: whenever an
  endpoint $\PointerC$ is received, the continuation process cannot
  use any endpoint other than $\PointerC$ and the one from which
  $\PointerC$ was received.
}


\section{Conclusion and future work}
\label{sec:conclusion}

In~\cite{BonoMessaPadovani11} we have formalized \MiniSing{}, a core
language of software isolated processes that communicate through
copyless message passing. Well-typed processes are shown to be free
from faults, leaks, and communication errors.
In the present paper we have extended the type system of \MiniSing{}
with bounded polymorphism, on the lines on what has been done for
session types in~\cite{Gay08}. Bounded polymorphism increases the
expressiveness of types and improves modularity and reusability of
components. In our setting, where resources -- and endpoints in
particular -- are \emph{linear}, we claim that bounded polymorphism is
even more useful in order to avoid the loss of type information that
occurs when endpoints are delegated and therefore exit the scope of
the sender process (Section~\ref{sec:example}).
We have shown that the \emph{finite-weight} restriction we introduced
in~\cite{BonoMessaPadovani11} scales smoothly to the richer type
language, despite the fact that polymorphism augments the critical
situations in which a leak may occur (recursive types are no longer
necessary to find apparently well-typed processes that leak memory, as
shown in Section~\ref{sec:introduction}). This is mostly due to a nice
property of weights (Proposition~\ref{prop:subt_weight}).
Unlike~\cite{BonoMessaPadovani11}, we have omitted cells and open cell
types, basically because they can be encoded thanks to the increased
expressiveness given by (bounded) polymorphism (Examples~\ref{ex:cell}
and~\ref{ex:cell_types}).

\REVISION{With respect to \Sing{}, our model still differs in a number
  of ways: first of all, \PolySing{} processes are terms of a process
  algebra, while \Sing{} is an imperative programming language similar
  to \CSharp{}. As a consequence, there is no direct mapping of
  \Sing{} programs onto \PolySing{} processes, even though we claim
  that our calculus captures all of the peculiar features of \Sing{},
  namely the explicit memory management (with respect to the exchange
  heap), the controlled ownership of memory allocated in the exchange
  heap, and the contract-based communication primitives.
  Second, in~\cite{BonoMessaPadovani11} and in the present paper we do
  not deal with mutable record structures nor with mutable arrays but
  only with mutable cells. The extension to the general framework is
  straightforward and does not pose new technical problems.
  Finally, the fact that we do not take into account non-linear values
  is indeed a limitation of the model presented here, but we plan to
  extend it in this direction as described next.}

We envision two developments for this work. The first one is to enrich
the type system with non-linear types, those denoting resources (such
as permanent services) that can be shared among processes. Even though
it is plausible to think that the technical details are relatively
easy to work out (non-linear types are supported by~\cite{Gay08} and
in most other session type theories) we see this enhancement as a
necessary step for \PolySing{} to be a useful model of Singularity
processes.
The second development, which looks much more challenging, is the
definition of an algorithm for deciding subtyping
(Definition~\ref{def:subt}). As observed in~\cite{Gay08}, bounded
polymorphic session types share many properties with the type language
in the system
\Fsub~\cite{CardelliMartiniMitchellScedrov94}. Therefore, while it is
reasonable to expect that properties and algorithms for extensions of
\Fsub with recursive types~\cite{ColazzoGhelli05} carry over to our
setting, the exact details may vary.
%


\REVISION{\paragraph{Acknowledgments.} We are grateful to the
  anonymous referees for their detailed comments and reviews, and to
  the organizers of the ICE workshop for setting up the interactive
  reviewing process.}

\bibliographystyle{eptcs}
\bibliography{main}

\iflong
\appendix
\input{proofs}
\fi

\end{document}